\documentclass{emulateapj}

\usepackage{epsfig}
\slugcomment{ApJ, accepted}

\shorttitle{Arm \& Interam Star Formation in Spiral Galaxies}
\shortauthors{Foyle et al.}

\begin{document}

\title{ Arm \& Interarm Star Formation in Spiral Galaxies}

\author{K.Foyle, H.-W.Rix, F. Walter}
\affil{Max-Planck-Institute f\"{u}r Astronomie, K\"{o}nigstuhl 17, 69117, Heidelberg, Germany}
\email{foyle@mpia.de}
\author{A. Leroy}
\affil{National Radio Astronomy Observatory, 520 Edgemont Rd., Charlottesville, 
VA 22903 USA}

\begin{abstract}
We investigate the relationship between spiral arms and star formation in the grand-design spirals NGC~5194 and NGC~628 and in the flocculent spiral NGC~6946. Filtered maps of near-IR (3.6$\mu$m) emission allow us to identify ``arm regions" that should correspond to regions of stellar mass density enhancements. The two grand-design spirals show a clear two-armed structure, while NGC 6946 is more complex. We examine these arm and interarm regions, looking at maps that trace recent star formation --- far-ultraviolet (GALEX NGS) and 24$\mu$m emission ({\em Spitzer} SINGS) --- and cold gas --- CO (HERACLES) and HI (THINGS). We find the star formation tracers and CO more concentrated in the spiral arms than the stellar 3.6$\mu$m flux.  If we define the spiral arms as the 25\% highest pixels in the filtered 3.6$\mu$m images, we find that the majority (60\%) of star formation tracers occurs in the interarm regions; this result persists qualitatively even when considering the potential impact of finite data resolution and diffuse interarm 24$\mu$m emission.  Even with a generous definition of the arms (45\% highest pixels), interarm regions still contribute at least 30\%\ to the integrated star formation rate tracers. We look for evidence that spiral arms trigger star or cloud formation using the ratios of star formation rate (SFR, traced by a combination of FUV and 24$\mu$m emission) to H$_2$ (traced by CO) and H$_2$ to HI.  Any enhancement of SFR / M(H$_2$) in the arm region is very small (less than 10\%) and the grand design spirals show no enhancement compared to the flocculent target. Arm regions do show a weak enhancement in H$_2$/HI compared to the interarm regions, but at a fixed gas surface density there is little clear enhancement in the H$_2$/HI ratio in the arm regions.Thus, it seems that spiral arms may only act to concentrate the gas to higher densities in the arms.
 \end{abstract}


\section{Introduction}
Optical images clearly reveal that spiral arms in present-day disk galaxies
harbour a concentration of young stars, implying that star formation rate
densities must be higher in the arm regions than elsewhere in the galaxy.
There have been many attempts to understand the connection between star
formation and spiral arms and, given the variety of spiral structures
observed, it is likely that more than one model may be required to explain
observations.  We can distinguish between two types of spiral structure.  In the first, the entire disk participates in the spiral pattern and is thus
well-defined in all bands so that it is not only associated with young, star
forming regions, but also the underlying mass density (Eskridge et al. 2002, Binney \& Tremaine 2008).  The second type has a spiral pattern that is not well-defined and is seen only in the optical bands, with little presence in redder bands commonly used to probe the stellar mass density (Elmegreen \& Elmegreen 1984,  Binney \& Tremaine 2008, Zibetti et al. 2009).  Typically one finds the former in so-called `grand design' spirals with two-arm symmetry and the latter in flocculent spirals with multiple, short, spiral segments.  

In the case of the grand design spirals, the underlying variations in the stellar mass density may simply lead to the reorganization of the ISM (e.g. Elmegreen 1995 and Elmgreen \& Elmgreen 1986).  The gas is drawn toward the mass enhancements, which define the arm areas and is retained for a longer time due to Coriolis forces.  High star formation rates in the arm areas may only be due to the higher gas densities and the star formation rate per unit gas mass, or star formation efficiency is the same throughout the disk.  However, spiral arms could conceivably do more: as first proposed by Roberts (1969)  and extended by others (e.g. Roberts et al. 1975, Gittins \& Clark, 2004), the spiral arm mass density enhancement could act to directly trigger star formation by a shock forming along the trailing edge of the spiral arm inside corotation when the relative velocity between the interstellar medium and the density wave is supersonic.   The shock compresses the gas, which leads to star formation and one would expect the star formation rate per unit gas mass to be higher in the spiral arms than the interarm regions.  We will refer to this last scenario as the `triggering model'. 

In the more flocculent spirals without a pattern in the stellar mass density, or only a very weak one (Thornley 1996, Kudo et al. 1997), much of the spiral pattern seen in optical images is a consequence of sheared, stochastic star formation (e.g. Gerola \& Seiden, 1979 and Elmegreen et al., 2003).  In this way, star formation causes the spiral pattern to emerge.  Stars form in small patches and are sheared into a spiral pattern by differential rotation.

Our ability to decipher which of these scenarios is most important is complicated by several factors. It is difficult to define the spiral arm regions and it is still unclear whether they are long-lived, quasi-stationary structures or short-lived, transient structures (Sellwood, 2010).  Moreover, the timescales for the stages of star formation are not well-known and we do not have direct measures of the star formation rates and star formation efficiencies.  

To assess empirically how spiral arms affect star formation, we aim here to provide two key pieces of information: first, the fraction of star formation that occurs in the arms, as opposed to the interarm regions.  Second, whether the star formation efficiency (SFE) is higher in the arms as opposed to the interarm regions.  Few studies have attempted to address the former and have focused on the latter.  The amount of star formation in the interarm region is very important in order to assess how relevant spiral arms are in the overall production of stars in galaxies.  If the fraction of star formation in the arms were modest, the effect of spiral arms on the net production of stars would still be small irrespective of any triggering.    Our sample includes three galaxies, two grand design and one more flocculent, allowing us to explore the range of possible models.  We first review some of the previous works, which have largely focused on grand design spirals and whether or not the spiral arms directly trigger star formation.

Elmegreen \& Elmegreen (1983) examined 34 spiral galaxies in the blue and near-infrared bands and found that the blueness of the arms was independent of the arm amplitude.  The triggering model would predict that higher arm strengths and hence shock strengths should lead to higher star formation rates and thus more young, blue stars.  The lack of such a correlation supports the reorganization model.  However, Seigar \& James (2002) used individual galaxy estimations of the spiral arm shock strengths and found a coupling with the H$\alpha$-based star formation rates (SFR).

The triggering model also predicts that grand design spirals should have higher star formation rates than their non-grand-design counterparts and that the properties of the arms including width and pitch angle should be correlated with Hubble Type.   Elmegreen \& Elmegreen (1986) found that the galaxy-averaged star formation rates determined from H$\alpha$ and UV fluxes showed no difference for galaxies with or without grand design spirals.  However, for M51, Vogel et al. (1988), found that the SFE in terms of H$\alpha$ and CO was higher in the arm region, but that only 25\% of the CO emission was found in the arms.  As a result of this small fraction of gas in the arms, even a strong enhancement in the on-arm SFE will weakly affect the integrated SFE of the galaxy.  Kennicutt \& Hodge (1982) and Seigar \& James (1998)  found that correlations between arm properties and Hubble type did not match the predictions of the triggering scenario. 

There have been many studies of the arm versus interarm star formation efficiency, particularly for M51.  Studies of M51 by Lord \& Young (1990) have found that the  star formation efficiency expressed as H$\alpha$/CO was higher in the arm region versus the interarm region.  Conversely, Garcia-Burillo et al. (1993) claim that arm-interarm contrasts of CO in M51 can be explained by orbit crowding and that triggering need not be invoked.    Rand \& Kulkarni (1990) studied giant molecular associations via CO measurements in the arm and interarm regions of M51 and found that these associations were found in both regions, but only  those in the arms were bound.  This suggests that the density wave may trigger the formation of molecular gas, but may not trigger star formation or enhance the SFE.  Koda et al. (2009) also detected molecular gas throughout the disk and the giant molecular associations were interpreted to be forming due to streaming motions as they approached the arms and were then fragmented due to shear as they left the arms.

Beyond M51, a handful of other galaxies have also been studied.  Knapen et al. (1996) found that the arm regions had SFEs three times higher than the interarm regions for NGC 4321 in terms of H$\alpha$/CO.  Cepa \& Beckman (1990) reported higher star formation efficiencies in the arm regions for NGC 6946 and NGC 628 using H$\alpha$ and HI maps.  It is important to note though that the SFE in terms of HI and CO (molecular gas) are likely to be quite different.  Leroy et al. 2008 (hereafter, L08) and Bigiel et al. 2008 (hereafter, B08) has found that the star formation rate is not correlated to the HI distribution but rather the molecular gas.  Thus, it is important to distinguish whether the SFE is associated with the total gas, HI or H$_{2}$.  When comparing these studies, it is also important to examine closely how the spiral arms are defined.  In some cases, dust lanes are used and in other cases optical or near-infrared images are used.  Clearly, defining the spiral arms using a tracer sensitive to regions of recent star formation may lead to erroneous measures of high SFE.  Thus, it is important to probe the underlying density enhancement when defining the spiral arms.

The recent work by L08 and B08 has shown that the star formation efficiency of H$_{2}$(SFE=SFR/M(H$_{2}$))  alone is constant to first order for nearby disk galaxies on a pixel-by-pixel basis.  However, they did not explore specifically whether the SFE might vary in the arm versus interarm regions.  In light of the recent increase in multiwavelength data for nearby galaxies from surveys including GALEX (Gil de Paz et al., 2007), SINGS (Kennicutt et al., 2003), THINGS (Walter et al., 2008) and HERACLES (Leroy et al., 2009), it is timely to explore how much and how efficiently star formation happens in the arm and interarm regions. 

Looking at the nearby galaxies with prominent spiral arms, we first focus on observable tracers of star formation and gas and ask what fraction of them lies near spiral arms.  We then use the tracers to estimate the SFR and SFE and examine if there are any differences in the arm and interarm regions.  We also examine whether the fraction of molecular gas, (M(H$_{2}$)/M(HI)), is enhanced in the arm and interarm regions in order to determine if the arms are triggering molecular gas formation.  In \S 2 we describe how the images are processed and how the spiral arms are defined.  In \S 3 we examine how concentrated the star formation and gas tracers are in the spiral arms relative to the stellar mass density and infer the amount of interarm star formation.  In \S4 we transform our observables into estimates of the star formation rate and star formation efficiency in the arm and interarm regions and compare these regions.  In \S5 we compare the fraction of molecular gas in the arm and interarm regions and specifically if it is enhanced relative to other gas of similar surface density. In \S6 we calculate the fraction of diffuse 24 $\mu$m emission for our sample.  We summarize our conclusions in \S 7.

\section{Analysis}
We chose three galaxies, two grand-design (NGC 628 and NGC 5194) and one more flocculent  (NGC 6946) on the basis of their proximity, orientation and multiband data.  These have coverage in GALEX (NUV+FUV) (Gil de Paz et al., 2007), THINGS (Walter et al., 2008), SINGS (Kennicutt et al., 2003) and IRAM 30 observations (Schuster et al. 2007, Leroy et al., 2009).   

In order to quantify the amount of star formation in the arm and interarm regions we focus on a series of observables: 24$\mu$m emission, which traces young, dust enshrouded stars; UV emission, which traces young, unobscured stars; CO, which traces molecular gas that is presumably organized into giant molecular clouds; and HI gas, which is presumably a mixture of warm and cold, atomic, diffuse, gas.  Each of these probes a different stage in the star formation process.  However, none of them uniquely define the star formation rate as other sources may contaminate or attenuate the emission.  The 24$\mu$m emission, especially in the interarm region, may also arise from diffuse emission not associated with recent star formation (cirrus 24$\mu$m emission) (Helou, 1986, Calzetti et al. 2007).  The FUV, on the other hand, may be attenuated by dust in the arm region (Kennicutt 1998, Calzetti et al. 2007).  Taken together, these two effects could potentially boost the relative amount of star formation in the interarm region.  Fortunately, CO measurements, which trace the molecular gas component, i.e. the fuel for star formation, can also be used to further probe the amount star formation in the interarm region, if we assume that the relations found by B08 and L08 indeed hold.  We first describe the steps taken to render the images for our analysis as well as how the arm and interarm regions are defined in the following subsections.   
\subsection{Images}
All of the images are aligned to the THINGS astrometric grid and degraded to a common resolution of 13$"$ FWHM, which is the resolution of the HERACLES CO images.  Before this degradation, we remove foreground stars from the far-UV, 24$\mu$m and 3.6$\mu$m images using their UV colour.  Pixels with an NUV-to-FUV intensity ratio between 9 and 25, depending on the galaxy, are blanked.  We also require that the cut pixels have values greater than 5$\sigma$ in the NUV map.  The companion of NGC 5194 is removed by-eye, and is beyond the radius considered in the analysis.  All images are eventually deprojected to face-on according to the values found in Walter et al., 2008 and listed in Table 1.

\begin{table}
\begin{center}
\caption{Sample Properties}
\begin{tabular}{ccccccc}
\hline\hline
Name & inc &  PA & R$_{in}$  & R$_{out}$  & R$_{in}$& R$_{out}$ \\
& [$^{\circ}$] & [$^{\circ}$] & [$''$] & [$''$] &  [kpc]  &  [kpc] \\
NGC 5194 & 20 & 172 & 20 & 95 & 0.8 & 3.7\\
NGC 628 & 7 & 20 & 30 & 76 & 1.1 & 2.7\\
NGC 6946 & 33 & 243 & 20 & 107 & 0.6 & 3.1\\
\hline

\end{tabular}
\tablecomments{The inclination and position angles used to deproject the galaxies and the inner and outer radii in $''$ and kpc defining the region of the analysis are listed.}
\end{center}
\end{table}

\subsubsection{UV, 24 $\mu$m \& SFR Maps}
As in L08 and B08,  we remove a residual background from the FUV and 24$\mu$m images, measured as a median value in an off-galaxy box.  Stars are removed using the NUV-to-FUV ratio and the UV images are corrected for galactic extinction (Schlegel et al. 1998) using the E(B-V) values listed in NED, which are 0.07, 0.035 and 0.342 for NGC 628, NGC 5194 and NGC 6946, respectively.    The UV and 24$\mu$m images are then combined to produce SFR maps (see Appendix of L08):

\begin{equation}
\Sigma_{SFR}=(8.1 \times 10^{-2} I_{FUV} +3.2 \times 10^{-3} I_{24}),
\end{equation}
where $\Sigma_{SFR}$ has units of M$_{\sun}$ kpc$^{-2}$ yr$^{-1}$ and the FUV and 24 $\mu$m intensity are each in MJy ster$^{-1}$.

\subsubsection{HI, CO \& Gas Maps}
We use 21 cm line emission from THINGS to trace the atomic gas. We convert
the integrated intensity to a surface density and include a factor of 1.36 to
account for helium, following L08. 
We use integrated CO $J=2 \rightarrow 1$ intensity maps from HERACLES (Leroy et al., 2009) to estimate the distribution of H$_2$. For M51 this is a reprocessing of the maps presented by Schuster et al., 2007 and  Hitschfeld et al., 2009 (the reprocessing does not significantly alter the map). To estimate the surface density of H$_2$ from CO we use a constant conversion factor.  As in L08 we adopt:
\begin{equation}
\Sigma_{H_{2}} [M_{\sun} pc^{-2}] =5.5 I_{CO} (2 \rightarrow 1) [K km s^{-1}]
\end{equation}

The maps are deprojected and added together to form total gas density maps.    

The CO-to-H$_2$ conversion factor is a source of uncertainty. It should in principle be a function of (at least) metallicity, radiation field, density, and temperature and at least some of these conditions may change between the arm and interarm regions. However, direct evidence for conversion factor variations in our targets is mixed and often contradictory (e.g. Garcia-Burillo et al., 1993)  For a detailed discussion see Schinnerer et al., 2010. The fact that the studies of L08 and B08 showed striking agreement of different galaxies in the `Schmidt Law' plot (plotting SFR surface densities vs. gas surface densities) provides further confidence that the X$_{\rm CO}$ conversion factor is roughly constant for the systems studied here. Thus, we use a constant X$_{\rm CO}$ in our analysis.

\subsection{Defining Spiral Arms}
It is clear that we should define the arm and interarm regions, in a way that is least biased by the young stellar population.  Ideally, we wish to define the spiral arms using the stellar mass density, or at least the old stellar population. Near-infrared images have commonly been used for this purpose (e.g. Elmegreen \& Elmegreen 1984, Rix \& Zaritsky 1995, Seigar \& James, 1998, Grosbol et al., 2004 and Kendall et al. 2008). We use the 3.6$\mu$m images from the IRAC instrument on Spitzer (Kennicutt et al. 2003) to trace the underlying old stellar population. In this band most of the emission is due to old stars although there is some patchy contamination from hot dust and PAH features (e.g. Kendall et al. 2008).  Zibetti et al. 2009 provide a much improved algorithm for mapping out the stellar mass density, which makes use of multi-band images.  However, here we do not require the exact amplitude of stellar mass density variations, but only the location of mass density enhancements to define the {\it location} of spiral arms.   

\begin{figure}
\includegraphics[scale=0.4]{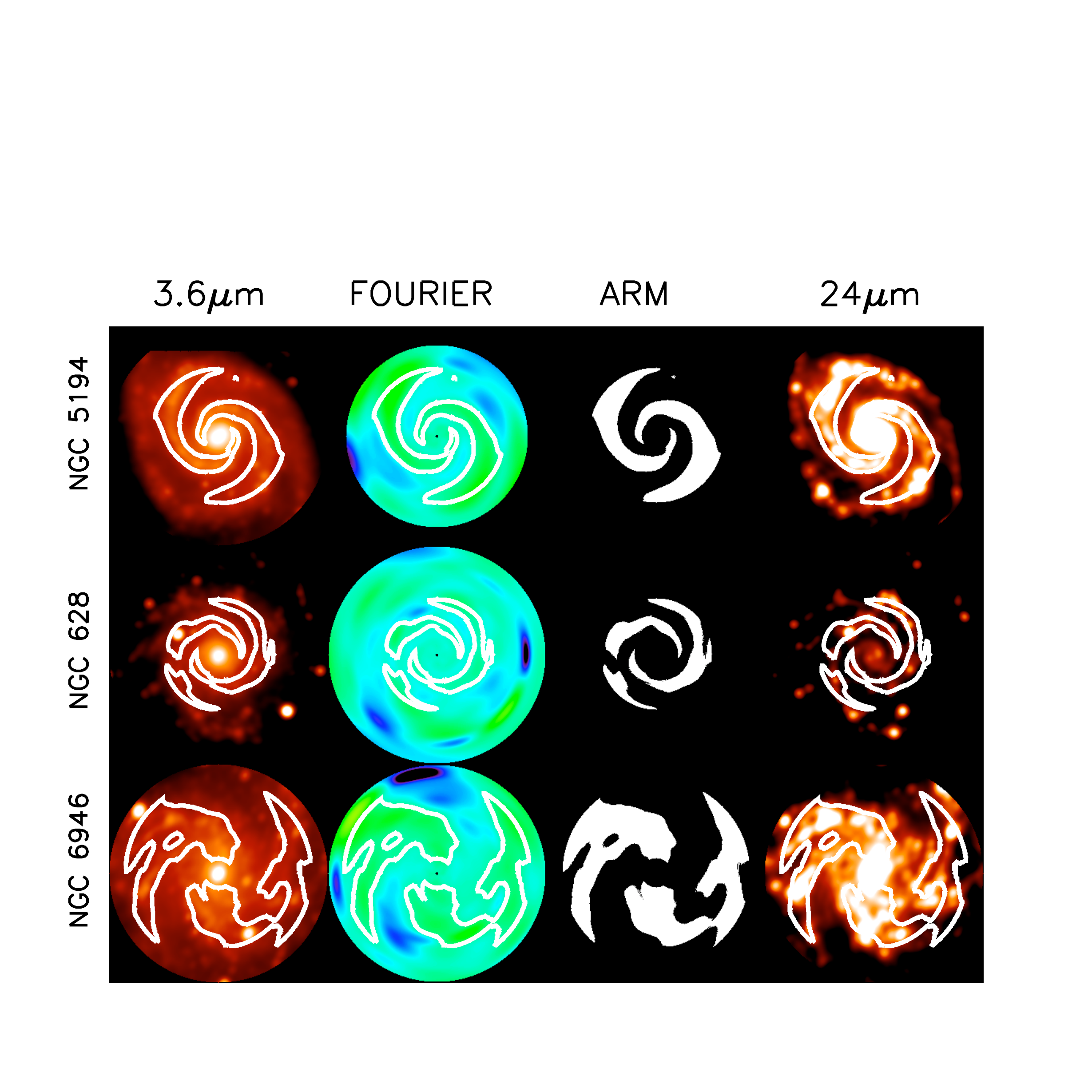}
\caption{Identification of arm and interarm regions.  The 3.6$\mu$m image (left) and the Fourier reconstructed m=6 image divided by the m=0 component (second from left) with contours showing the arm (white) regions for NGC 5194, NGC 628 and NGC 6946.  The arm (second from right) regions are defined by the 45\% highest pixels in radial annuli of the m=6 component divided by the m=0 component of the Fourier reconstructed 3.6$\mu$m image. The 24$\mu$m image (right) is shown with the arm contours of overlaid (white).}
\label{mask}
\end{figure}

Foreground stars in the processed 3.6$\mu$m images are removed according to the UV color cut described above and the images are deprojected according to the values in Table 1.  In order to make the 3.6$\mu$m surface brightness variations a better approximation to the {\it location} of stellar mass density enhancements, we apply some spatial filtering.  The 3.6$\mu$m images are first median filtered over 20 pixels ($\approx$1 kpc), to remove bright spots and features.  The filtered image is then Fourier-decomposed in $\phi$ with radial bins that overlap to obtain a version of the image that consists of the m=6 component divided by the m=0 component.  This spatially filters the images, which are shown in Figure~\ref{mask}. We note that the grand-design spirals, NGC 5194 and NGC 628, have spiral arms that were well-defined using only the m=4 component divided by the m=0 component and that moving up to m=6 leads to little, if any difference.  However, in the more flocculent spiral, NGC 6946, the 3.6$\mu$m  structure is more complex, requiring an extension to m=6.

The inner bulge regions, where no spiral arms are evident, are masked out in each galaxy.  The outer limit is set by the area over which the CO maps detect emission, or, in the case of NGC 5194, when the arms become too tightly wound for accurate definition by the mask.  In all three cases the analysis does not extend to the outer regions of the galaxies.  We extend to 0.3 or 0.4$r_{25}$ depending on the galaxy, where most of the star formation and luminosity is found.  The inner and outer limits for each galaxy are listed in Table 1.  The image is then divided into radial annuli each of 7.5$''$ width.  We chose a width below the resolution of the image in order to ensure overlapping radial bins, which produce a more continuous spiral arm structure in the masks.   In each annulus, the `arm region' is defined to be the area covered by a certain percentage of the highest-value pixels (e.g., the brightest 30\%). In a similar fashion, the `interarm region' is defined as the area covered by the same percentage of the lowest-value pixels. We will vary the exact percentage used to define these regions over the range 10--50\%\ to ensure conclusions robust to the precise arm definition. We refer to the percentage used to define the arms as the `arm pixel fraction'.  Once both the arm and interarm regions each consist of 50\% of the pixels, the entire surface area is covered.

Figure~\ref{mask} shows the 3.6$\mu$m image (far left) and the Fourier reconstructed m=6 image divided by the m=0 (second from left)  with contours overlaid showing the arm regions (white).  The spiral arm masks (second from right) for our sample where the arm regions are defined from the 45\% highest pixels per radial bin are also shown as is the 24$\mu$m image with arm contours overlaid for comparison.

\section{Star Formation \& Gas Tracers in the Arm and Interarm Regions}

\begin{figure*}
\includegraphics[scale=0.8]{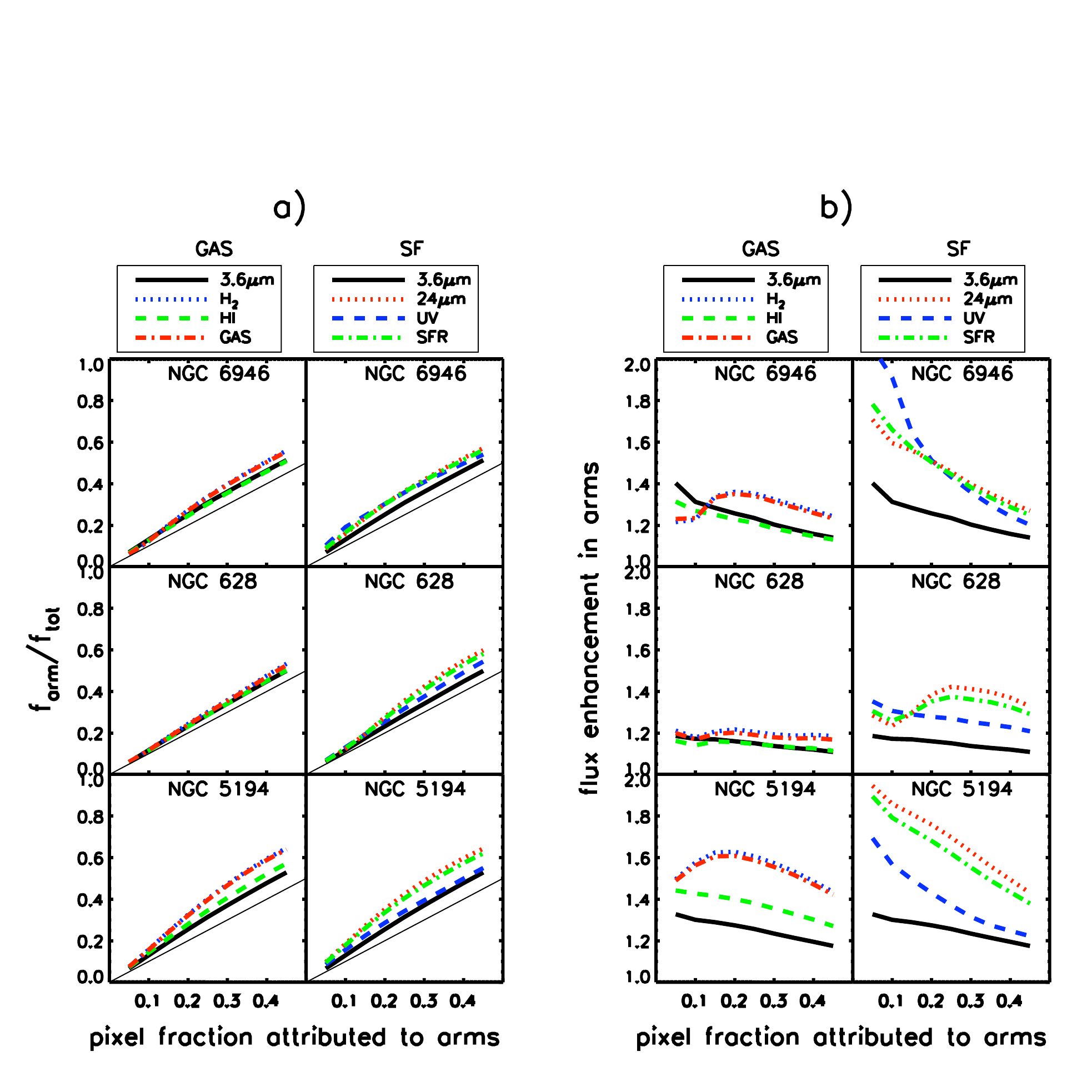}
\caption{Concentration of 3.6$\mu$m light, gas and SFR towards spiral arms. Fraction of total flux in each tracer in the arm regions (a) and the enhancement over a smooth distribution (b) (i.e. the curves in panel (a) divided by the thin black line).  The gas distributions are shown on the left
  columns of both panels and the star formation tracers are shown on the right columns for
  NGC 6946 (top), NGC 628 (middle) and NGC 5194 (bottom).   A
  uniformly distributed tracer is represented as a the thin black line in a);  in b), such a distribution would be a horizontal line at value 1.0.   The tracers are concentrated to the spiral arms, yet an appreciable fraction of the flux must lie in the interarm region (at least 30\% even when 45\% of the pixels, the last plotted pixel fraction in these plots, are enclosed in the spiral arms).  The 24$\mu$m emission is more concentrated to the spiral arms than the UV emission and the H$_{2}$ emission is more concentrated to the spiral arms than the HI emission.}
\label{six}
\end{figure*}

Having defined the arm and interarm regions, we now can assess what fraction of star formation tracers is found in the respective regions.   To do so, we measure the fraction of the total emission from gas and
  star formation tracers that are contained in the arm regions for the area considered (see Table 1 for inner and outer radii). We
  make this measurement for a variety of arm pixel fractions.  We examine the HI, H$_{2}$ (as traced by CO), total gas, FUV, 24 $\mu$m, SFR, stellar mass surface density.   The plots in Figure~\ref{six} show the fraction of the overall emission found in the arm regions, as a function of the arm pixel fraction.

Figure~\ref{six} a) shows the flux fraction for different tracers that
occur in the arm region as a function of the pixel fraction used to define the
arm region. To emphasize the difference between the curves, we have divided by the expectation of a spatially uniform distribution and shown them in b). The 3.6$\mu$m image is included
in all panels for ease of comparison.  The thin black line in a) denotes
what would be expected for an azimuthally uniform distribution.    In b), an azimuthally uniform tracer would be a horizontal line (flux enhancement = 1).  For ease of comparison in our discussion we use a fiducial pixel fraction of 45\% to define the arms.  

Figure~\ref{six} shows that all tracers are much more concentrated to the spiral arms than a uniform distribution (i.e. all values are above unity in panel b)).  For the 3.6$\mu$m image this is by construction, as the wavelength was used to define the arm regions.  However, Figure~\ref{six} shows that all the tracers of star formation are even more concentrated.  Once a sufficient fraction of pixels in the arm region is enclosed, the 24$\mu$m emission is more concentrated to the spiral arms than the UV emission.  UV emission is presumably less concentrated in the arms than 24$\mu$m, because dust extinction suppresses UV emission in the arms and because UV is a more `long-lived' SFR tracer than 24$\mu$m.  The effect is most pronounced for the two grand design spirals, NGC 5194 and NGC 628.   Similarly, the H$_{2}$ (CO emission) is more concentrated to the spiral arms than the HI.  The HI is the least concentrated to the arms. 

Figure 2 a) shows that, even for high arm pixel fractions, at least 30\% of emission from star formation tracers are found in the interarm regions for galaxies which were chosen for the prominence of their spiral arms.  We also note that both NGC 628 and NGC 5194 are grand design spirals, which should, in principle, exhibit the highest fractions of star formation in the arms, if grand design spirals have the strongest arms and shocks.  Assuming that the combination of 24$\mu$m and UV emission is an accurate description of the star formation rate, these plots also show that at least 30\% of the star formation tracers are in the interarm region, even if the definition of the spiral arm encompasses 45\% of the total area in a radial bin.  Thus, while spiral arms are important sites of star formation, star formation occurs throughout the disk in the interarm regions as well. 

One concern is that the interarm SFR tracer emission is due to stars which have drifted from the spiral arms.  If this were the case one would expect offsets between the different star formation tracers, which reflect the timescales in the star formation process.  In these three galaxies we found no evidence for such offsets between any of the tracers considered here.  Previous works, including Tamburro et al. 2008, found offsets to be very small between HI and 24$\mu$m (five degrees), which imply timescales between these two stages of less than 4 Myr (Tamburro et al., 2008).  Given this, any drifted emission would be well within our broad definition of spiral arms, which encompasses and ever increasing area.  However, the UV emission traces not only the current star formation, but also the recent star formation.   The timescales are much longer ($\sim$ 100 Myr) and it is likely that the interarm emission was produced in the arms and has drifted downstream ({\it e.g.} Calzetti et al. 2005).  Indeed, the UV emission is less concentrated to the arms than the 24$\mu$m emission.  Our tracers, especially the 24$\mu$m emission, may also include diffuse emission in the interarm regions, which may contaminate our star formation indice.  Thus, we stress that our results hold for the star formation tracers.

\begin{figure}
\includegraphics[scale=0.4]{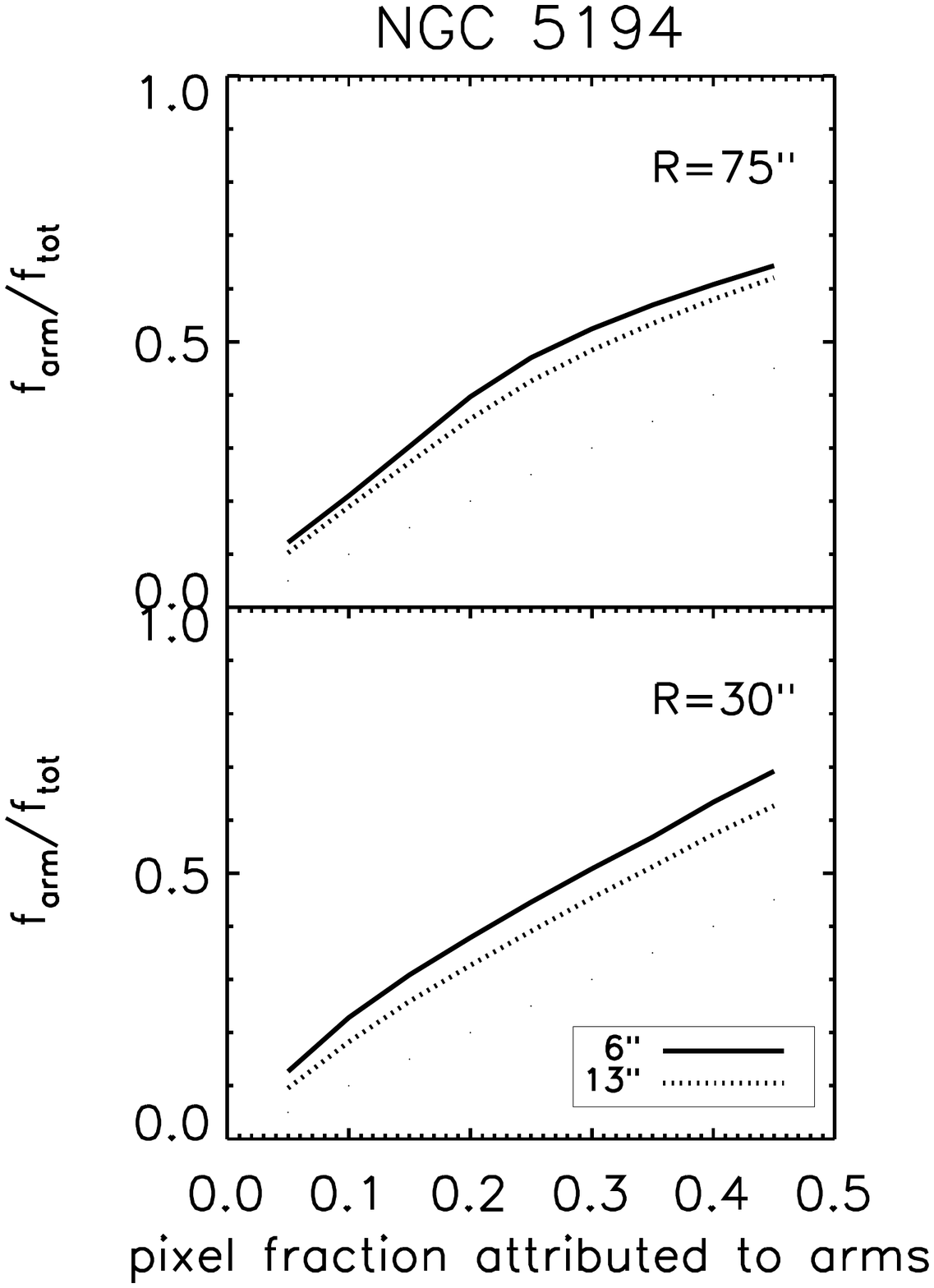}
\caption{The effect of spatial resolution on our estimates of interarm star formation tracers.   We use our estimate of the star formation rate based on UV and 24$\mu$m image for NGC 5194 at a resolution of 6$''$ and our common image resolution 13$''$.  We measure the fraction of flux attributed to the arms to the total versus the fraction of pixels enclosed in the maps.  At small radii (bottom), the lower resolutions bleed into the interarm area ($\approx$ 1kpc).  At larger radii (top), the effect quickly weakens once a sufficient number of pixels are part of the arm definition ($\approx$ 2.5 kpc). In both cases the effect is less than 10\%.}
\label{res}
\end{figure}

Another concern is that given the resolution of 13$''$, some flux physically arising in the arm region, may cause the arm flux to infiltrate the interarm flux.  In Figure~\ref{res}  we examine how resolution may affect our results for one case (NGC 5194).  We apply our analysis to our star formation rate maps, which are a combination of UV and 24$\mu$m images.  We consider star formation rate maps at  6$''$ and 13$''$ resolution to examine the effect of resolution. We find that particularly at small radii, resolution effects may lead to an overestimate of the star formation by as much as 10\%, depending on the choice of arm mask; at larger radii, the effect rapidly diminishes.  Due to this, for most comparisons we choose a pixel fraction of 45\% to define the arms.  Unless otherwise stated, one can assume this.  We also examined deconvolved star formation rate maps made by deconvolving each of the 24$\mu$m and UV emission maps with an estimate of their respective point spread function.  The deconvolved maps were then combined to form a deconvolved star formation rate map.  Even in this case there was still considerable emission in the interarm regions and the effect was at the 10\% level.  

\begin{figure}
\includegraphics[scale=0.3]{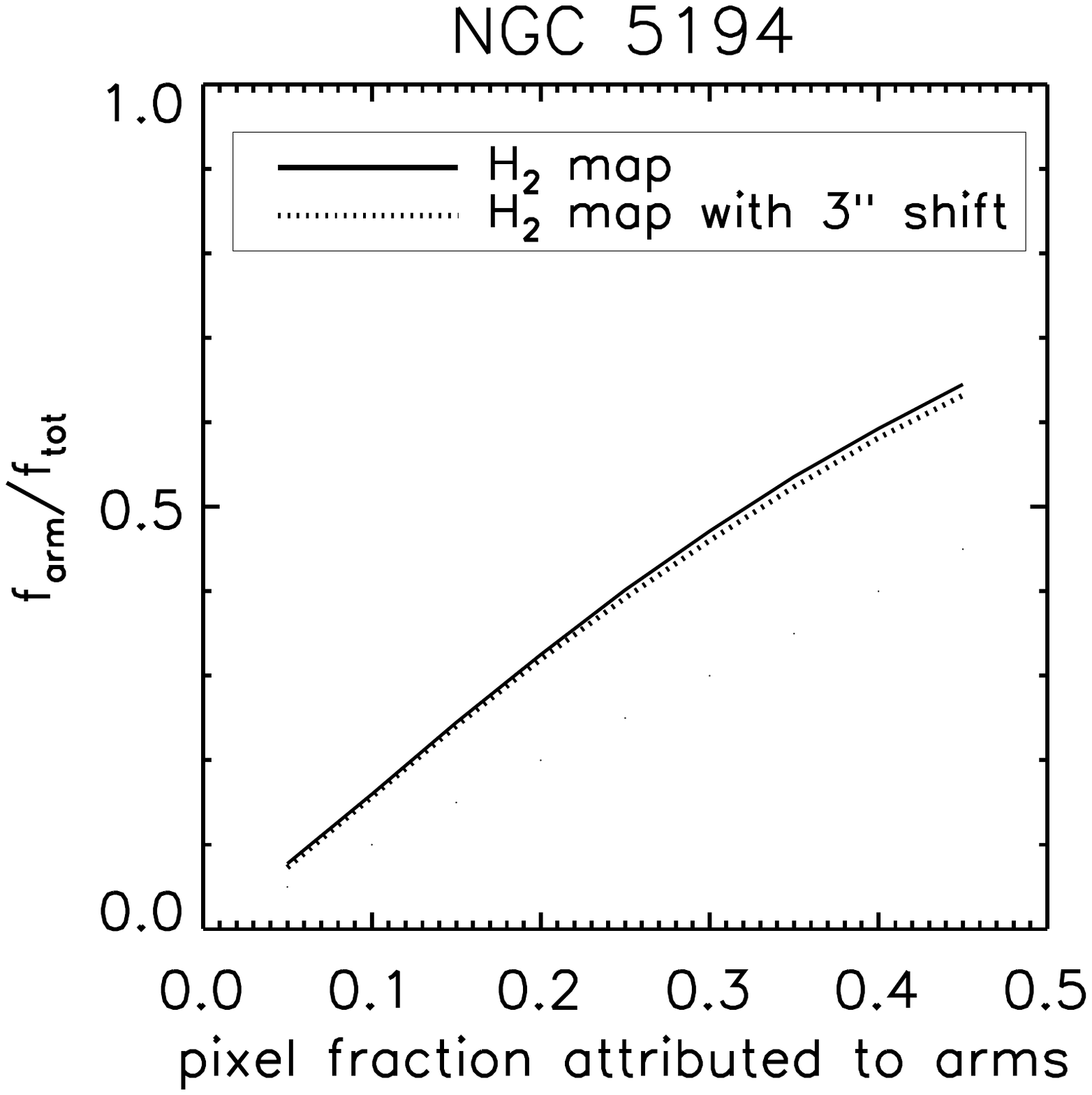}
\caption{The effect of finite pointing errors on H$_{2}$ maps.   We compare the results for the original H$_{2}$ map for  NGC 5194 with one that has been shifted by 3$"$ in both x and y.  We find that the effects of pointing errors are small and less than resolution effects.}
\label{pointing}
\end{figure}

We also examine the effects of possible pointing errors in the CO maps.  The HERACLES maps have been found to have offsets from BIMA SONG maps by less than $\pm$2$''$ in RA and DEC.  In Figure~\ref{pointing} we compare the results of a CO map shifted by 3$''$ in both x and y to the original map.  We find that the differences are minor and certainly less than those incurred by the resolution.

\section{The Star Formation Efficiency in the Arm and Interarm Regions}
Having addressed the relative fractions of arm and interarm star formation, we
now turn to exploring whether the SFE of H$_{2}$  differs in the arm and interarm regions.  L08 found that the SFE of H$_{2}$ of spirals was roughly constant with a median of 5.25 $\pm$ 2.5 $\times$ 10$^{-10}$ yr$^{-1}$.  Furthermore, L08 found no trends of the SFE with any macroscopic properties considered including, radial position, gas and stellar surface density, orbital timescale, gas pressure and the slope of the rotation curve.  Previous studies by Lord \&Young(1990), Cepa \& Beckman (1990), Seigar \& James (2002), etc. found evidence for triggering in spiral arms based on color differences, enhanced star formation rates or star formation efficiencies.  If spiral arms were shocking the molecular gas to produce stars with increased efficiency, then one would expect an increase of the SFE of H$_{2}$ in the arm region in comparison to the interarm region.  As L08 did not explore possible differences in the SFE(H$_{2}$) in these regions, we examine this question here.

\begin{figure}
\includegraphics[scale=0.6]{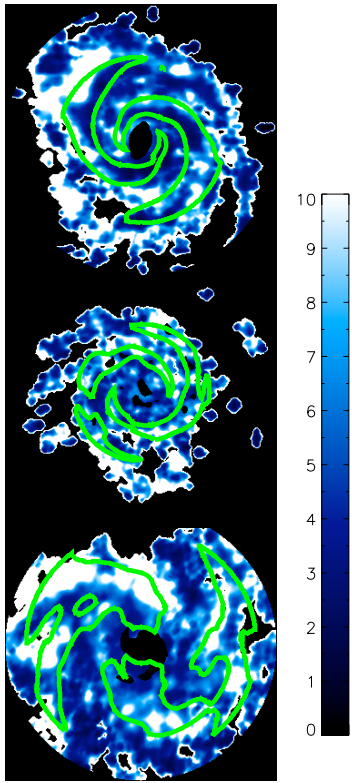}
\caption{The SFE for H$_{2}$ for NGC 5194 (top), NGC 628 (middle) and NGC 6946 (bottom).  Regions of high SFE are shown in white and low SFE are shown in blue.  The green contours show the spiral arm regions defined by 45\% of pixels. Pixels where the H$_{2}$ maps had values less than 4$\sigma$ were blanked. The colorbar has units of $10^{-10}$ yr$^{-1}$. }
\label{im}
\end{figure}

We first created molecular SFE maps using our SFR maps and H$_{2}$ maps for our sample (see Figure~\ref{im}, where we have blanked pixels where the H$_{2}$ map has values less than 4$\sigma$).   For display purposes, we have increased the contrast of these images as much as possible and have overlaid in green the spiral arm regions as defined by 45\% of pixels.  It is important to note that in the inner regions the SFR is dominated by the contribution from 24 $\mu$m emission and even in the outer parts it is dominate for M51.   Since we have restricted our analysis to the inner regions of the galaxies (see Table 1), our SFE is mostly determined from the CO and 24$\mu$m emission.    The SFE maps in Figure~\ref{im} exhibit no obvious spiral pattern.

\begin{figure*}
\centering
\includegraphics[scale=0.5]{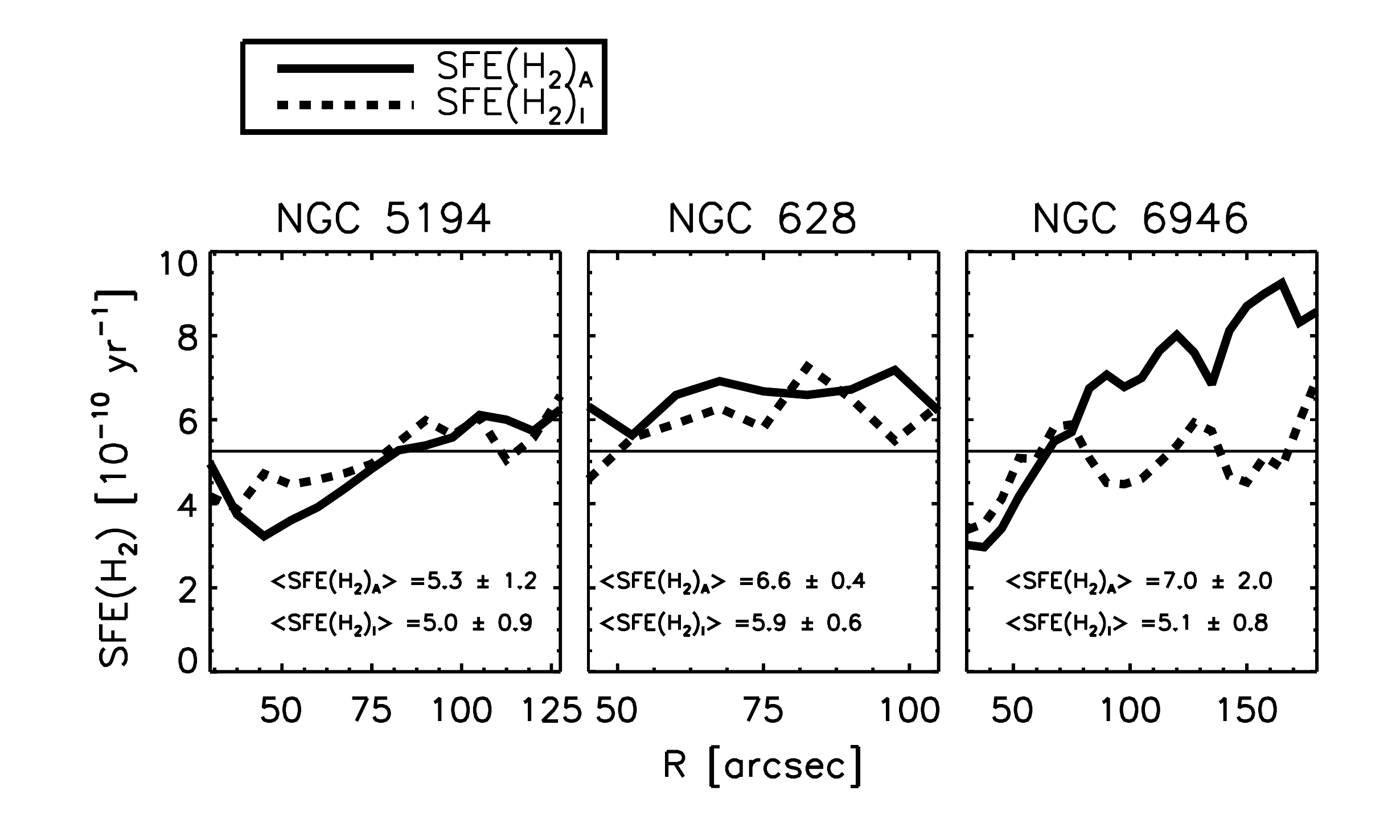}
\caption{The median value of the molecular SFE at different radii in the arm (solid) and interam (dashed) region for NGC 5194 (left), NGC 628 (middle) and NGC 6946 (right).  The median values over the whole galaxy are listed at the bottom of the figure.  The arm and interarm regions were chosen using the masks enclosing 45\% of pixels. The line shows the median value found by L08. }
\label{sfrco}
\end{figure*}

Figure~\ref{sfrco} compares the SFE(H$_{2}$) in the arm and interam regions for our sample.  We compare the median values of the SFE(H$_{2}$) at different radial annuli for the arm and interarm region.  For NGC 628 and NGC 5194, the variation is very small between the arm and interarm regions and is certainly less than the variation across the radial annuli.  For NGC 6946, the arm region has a higher SFE than the interarm region, particularly at larger radii (less than a factor of 2).  For the three galaxies there is a suggestive trend for an increase in the SFE with radius.  However, in the larger sample of L08 no trend with radius was found and the trend found here is small.

\begin{figure*}
\centering
\includegraphics[scale=0.5]{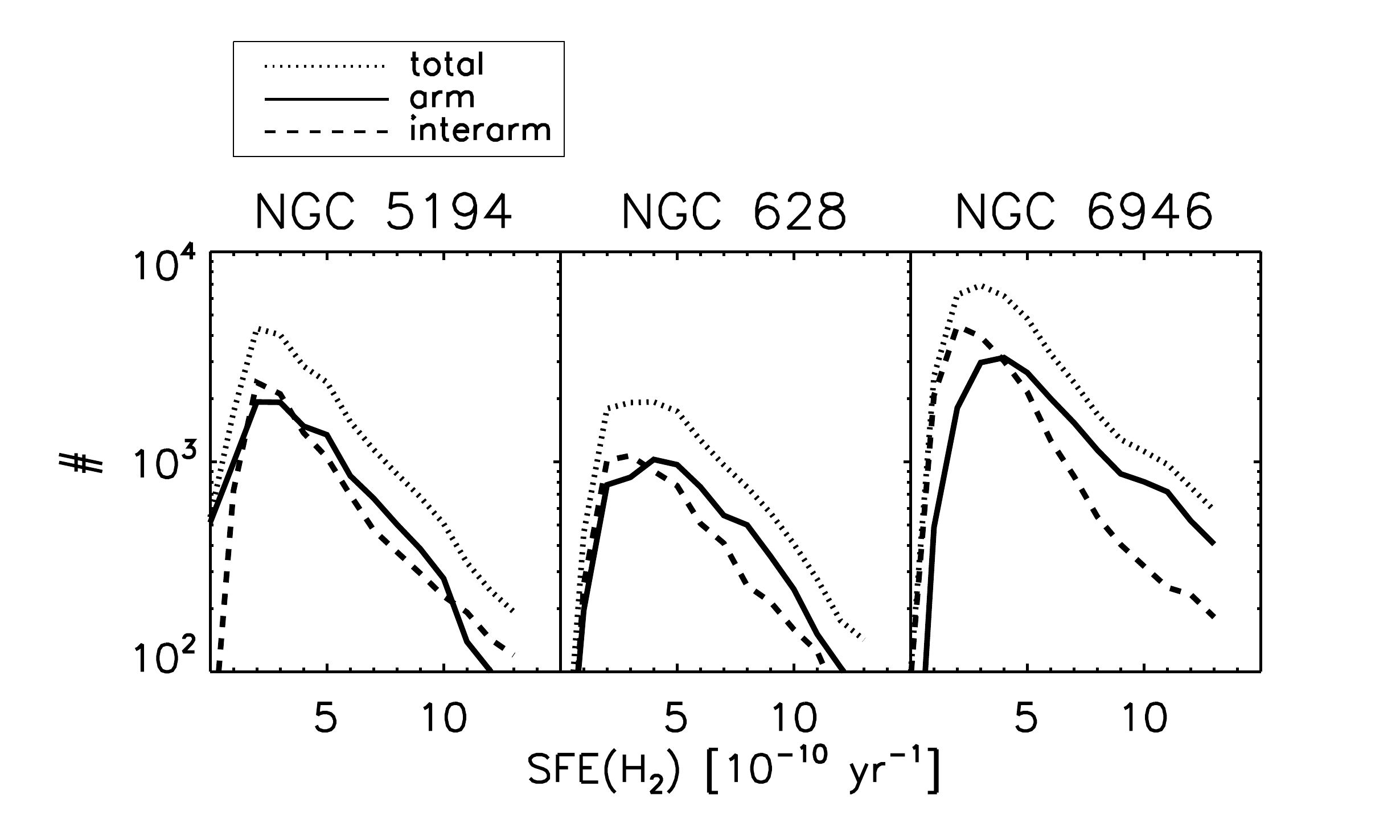}
\caption{Distribution of the pixel-by-pixel molecular SFEs for the total image
  (dotted), arm (solid) and interarm (dashed) regions for NGC 5194 (left), NGC
  628 (middle) and NGC 6946 (right).   The arm and interarm regions were
  chosen using the masks enclosing 45\% of pixels. We see that NGC 6946 shows
  an excess of higher SFE pixels in the arm region.}
\label{hist}
\end{figure*}

In Figure~\ref{hist} we show histograms of the pixel values in the total image (dotted), arm (solid) and interarm (dashed) regions.  Once again, we find that there is little difference between the arm and interarm region for NGC 5194 and only a slight difference for NGC 628.  However, NGC 6946 shows an excess of  higher SFE(H$_{2}$) pixels in the arm region (i.e., there are 33\% more pixels with a SFE value higher than 6 $\times$ 10$^{-10}$ yr$^{-1}$ in the arm regions versus the interarm regions).

One would expect that the grand design spirals would show the highest SFE in
the arms as opposed to flocculent galaxies, as here the spiral shocks should
be strongest.  It is interesting then that NGC 6946, the most flocculent galaxy in this study, seems to show a higher
SFE in the arms.  While the source of this is not clear, one possible
explanation is that our spiral arm definition is not probing the underlying
density enhancement for this galaxy, since it is very weak.  Instead the spiral arm
mask has isolated regions of high star formation.  Looking at the spiral arm
masks (see Figure~\ref{mask}), it is clear that the spiral arm structure is
much more complex than the grand design structures of NGC 628 and NGC 5194.
If the arms were defined based on young, recent star forming regions, then it would be biased towards high SFRs and hence show seemingly higher SFE in the arm regions for NGC 6946.  In order
to be able to compare the SFE in the arm and interarm regions it is essential
to have a definition of the arms that is not determined by episodes of recent star formation.  

Thus, at least for the two grand design spirals, NGC 628 and NGC 5194, we find
that the SFE based on the molecular gas component is not enhanced in the arm
region. Previous studies (e.g. Vogel et al. 1988, Lord \& Young, 1990, Cepa \& Beckman, 1990, Seigar \& James, 2002) claim to find an enhancement of the SFE in the arm region, but they used a different definition of the SFE than the SFR/H$_2$ than we use here. Most studies have looked at the SFE in terms of H$\alpha$ and HI.  As we saw in the previous section, the HI is far less concentrated in the spiral arms and is much closer to an even distribution.  Thus, it is not surprising to find an enhancement in the SFE in the arms, based on such a measure.  

Combined with our previous result that at least 30\% of the star
formation tracers  occurs in the interarm regions, we find no significant evidence for shock-triggered star formation by spiral arms.  The high star formation rates in the arms can be attibuted to the reorganization model.  We note that this does not imply that there are no dynamical effects at play.  It is possible that shock-triggering in the arms enhances star formation, but then shear flows act to inhibit star formation. Such canceling effects have been detected in highly barred systems (Zurita et al., 2004).  Momose et al. (2010) have recently shown that shear motions in the bar of NGC 4303 lead to a decreased star formation efficiency in the bar in comparison to the spiral arms.  However, while disk dynamics may affect star formation processes, our results show no evidence that star formation differs in an appreciable way between the arm and interarm regions on the spatial scales under consideration in this study (between 250-600 pc).

The fact that we find no evidence for shock-triggered star formation in spiral arms based on star formation rates does not preclude the possibility that the transition between atomic, neutral gas to molecular gas may be triggered by the passage of the spiral arm.  Indeed, the fact that the molecular gas is much more concentrated than the HI, suggests this to be the case.  Thus, we now examine if there are differences between the arm and interarm molecular gas fraction.

\section{Molecular Gas Fraction}
 Although we do not find an enhancement in the SFE of H$_2$, especially for the grand design spirals, Figure 2 does show an enhancement of H$_2$ relative to HI in the arms. This could be the result of molecular cloud formation triggered by spiral  arms, but this does not have to be the case. Arms represent concentrations of total (H$_2$+HI) gas. The fraction of gas in the molecular phase is a strong function of both the total gas surface density and the midplane average volume density (e.g., Blitz \& Rosolowsky 2006, L08, Krumholz et al., 2009). We now examine if there is evidence for an enhancement in the molecular gas in the arms and if this enhancement is independent of higher total gas surface densities in the arms.
 
\begin{figure*}
  \centering
\includegraphics[trim=60mm 20mm 30mm 150mm,scale=0.7]{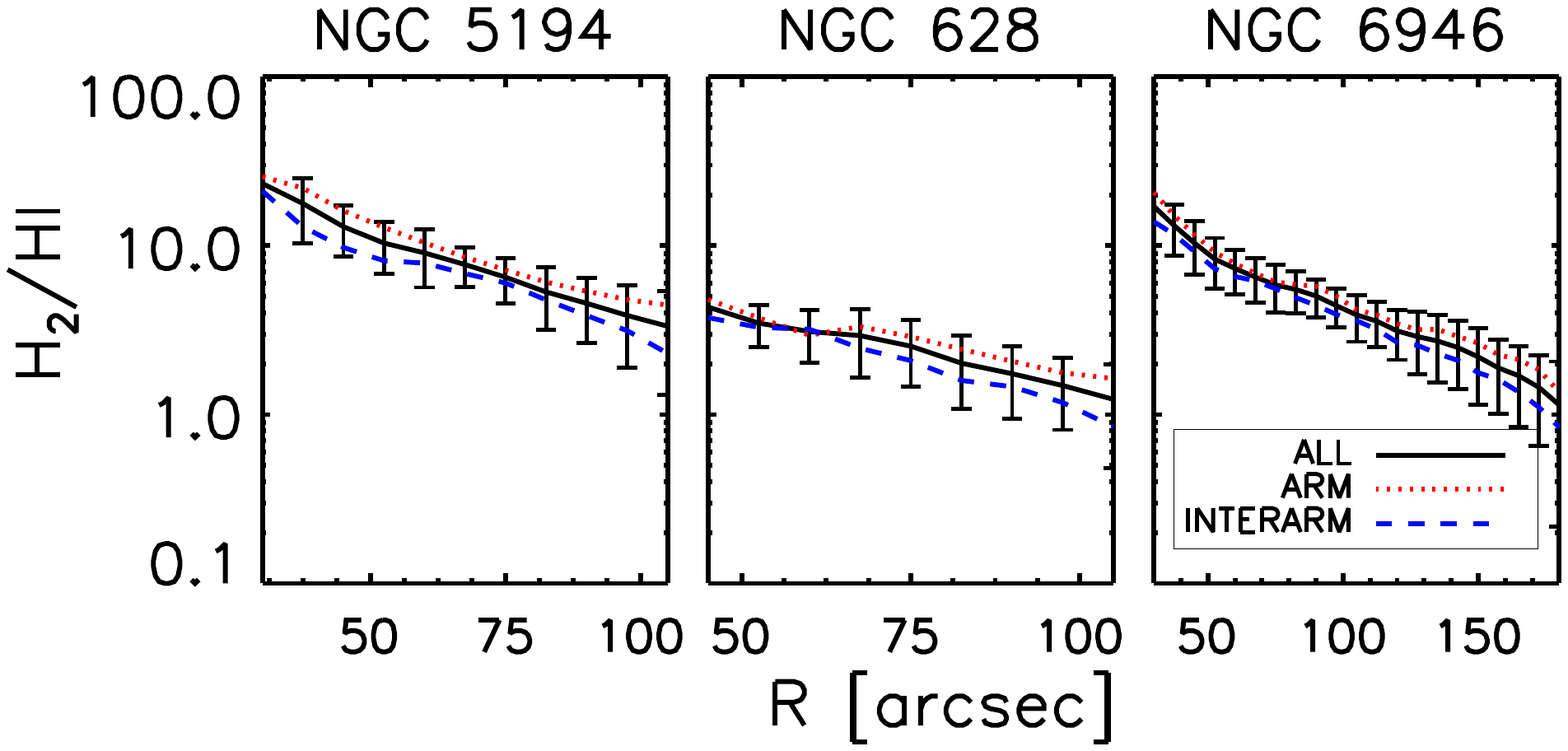}
\caption{Mean fraction of molecular gas (H$_{2}$/HI) in radial annuli for all pixels and their scatter (black) and arm (red) and interarm (blue) regions.  We find that any enhancement in the arm region is less than a factor of two. }
\label{gasr}
\end{figure*}

Figure~\ref{gasr} shows how the fraction of molecular gas, H$_{2}$/HI, varies with radius when considering all pixels (black), those attributed to the arms with the 45\% mask (red) and those attributed to the interarm regions (blue).  For each galaxy and radius, we find the median H$_{2}$/HI ratio in the arm regions is enhanced compared to both the interarm regions and overall trend. The magnitude of this enhancement is small, less than a factor of 2.

\begin{figure*}
  \centering
\includegraphics[trim=10mm 20mm 60mm 100mm,scale=0.7]{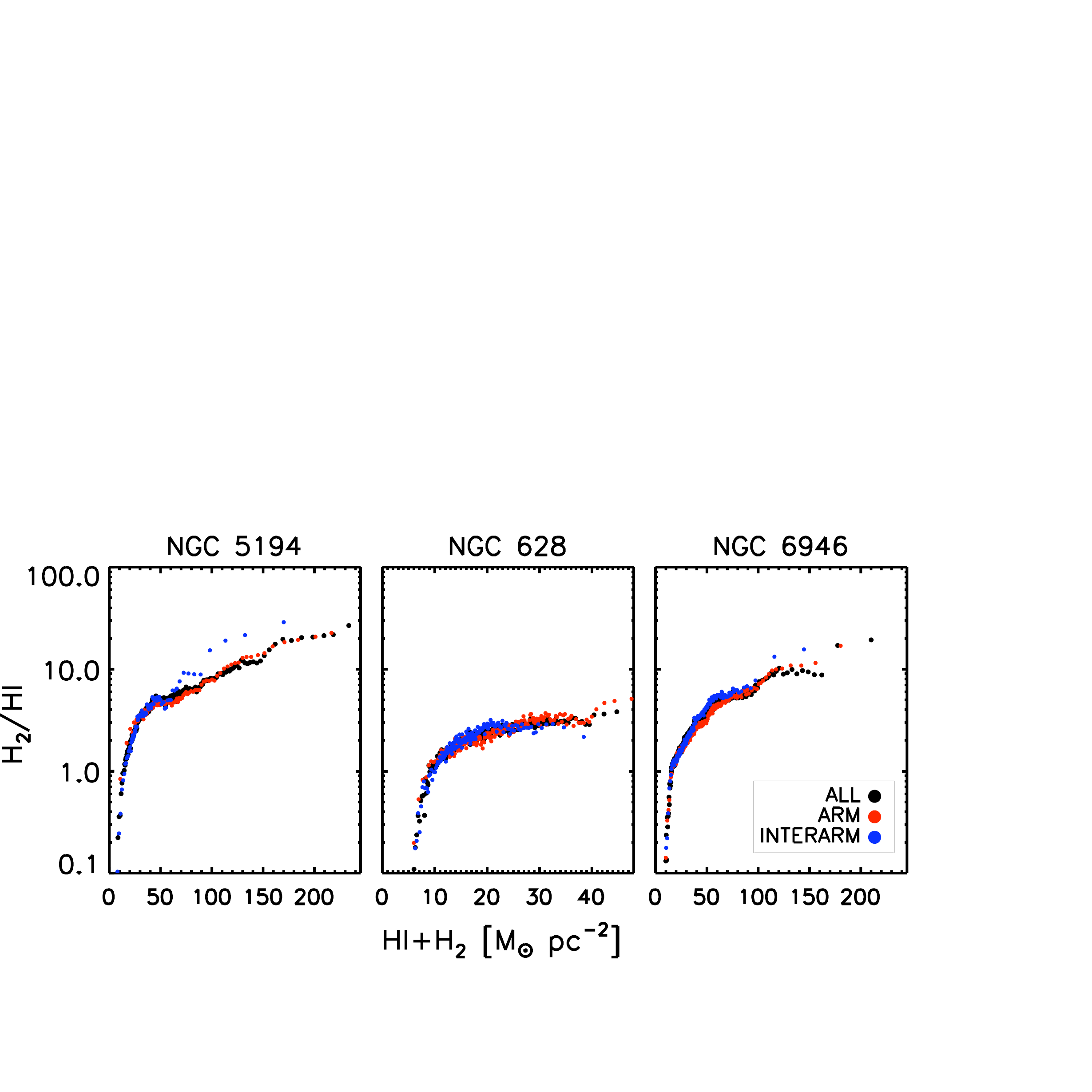}
\caption{Molecular gas (H$_{2}$/HI) as function of the total gas surface density for all pixels (black) and arm (red) and interarm (blue) regions.  In all three cases, there is no enhancement of the molecular gas fraction in the arms versus the interarm regions in terms of the total gas suface density. }
\label{gas}
\end{figure*}

Is this mild enhancement in the H$_{2}$/HI ratio the result of shock-triggered molecular cloud formation or simply the enhancement of the local gas content?  Figure~\ref{gas} shows the fraction of molecular gas, but this time in terms of the total gas surface density.  We see that the arm regions have molecular gas fractions that extend up to very high total gas fractions and that the interarm regions have molecular gas fractions only at the lower end of the total gas surface density.  However, the two overlap and there is no obvious enhancement of the molecular gas fraction in the arms for a given total gas surface density.

Thus, arms appear to concentrate gas to higher surface, and presumably volume, densities.   There is not strong evidence that arms trigger the formation of H$_{2}$, though.  Moreover, at a given gas surface density, the molecular gas fraction in the arm and interarm regions is about the same.  Combined with our SFE results, this suggests triggering by arms is not critical to the main star formation processes. Arms may drive the formation of molecular gas by bringing the total cold gas to high surface densities, but we do not see clear evidence that spiral shocks are contributing to form either clouds or stars.  However, we remind the reader that our study does not encompass the outer regions of these galaxies ({\i.e.} outside $\sim$ 0.35 $r_{25}$).  It is possible that in the outer regions, where the average gas density is too low to form stars, that the molecular gas fraction is enhanced due to the spiral arms.  Evidence for an enhanced star formation efficiency in the outer regions has been seen in some cases ({\it i.e.} Bush et al. 2010).

\section{Conclusions}
We have used three spiral galaxies (NGC 5194, NGC 628 and NGC 6946) to
determine the fraction of star formation and cold gas found in the interarm
regions of spiral galaxies.  We based our definition of the spiral arm areas on stellar mass density enhancements traced by 3.6$\mu$m images.  We find that at least 30\% of the emission of star formation tracers (far-UV and 24 $\mu$m images) must be located in the interarm region, showing that interarm star formation is significant even in grand design spirals.

We examined the star formation efficiency based on H$_{2}$ in the arm and interarm areas.  We confirmed the results of L08 that this quantity is constant on average and any enhancement in the arm areas is less than 10\% for the grand design spirals, NGC 628 and NGC 5194.  The flocculent spiral, NGC 6946, does show an enhancement of the SFE in the arm region, but this may be caused by an underlying weak spiral density wave, which has caused our spiral definition to be associated with isolated regions of high SFR.

We then explored whether the arms were triggering the formation of molecular gas by comparing the fraction of molecular gas in the arm and interarm regions.  The arms showed a higher molecular gas fraction, but this was found to be due to higher gas surface densities in the arms.

Taken together these results show that interarm star formation is significant and that the spiral arms gather the gas into regions of higher surface densities, which leads to an enhanced molecular fraction, but they do not ``shock trigger"  star formation nor molecular gas formation.  Thus, spiral arms act only to reorganize the material in the disk out of which stars form.  

\acknowledgements{K.Foyle acknowledges generous support from the Max Planck Society, International Max Planck Research School for Astronomy and Cosmic Physics at the University of Heidelberg and the National Science and Engineering Research Council of Canada.  We are very grateful to Karl Schuster and Carsten Kramer for providing the CO data on NGC 5194 and to Karl Schuster for his very helpful comments.  We also would like to thank the referee for making many useful suggestions on how to improve both the content and presentation of this work.}

\clearpage
\end{document}